\documentclass[superscriptaddress,nofootinbib,twocolumn]{revtex4-2}
\usepackage{amssymb}
\usepackage{amsmath}
\usepackage{mathdots}
\usepackage{slashed}
\usepackage{verbatim}

\newcommand{\be}{\begin{equation}}
\newcommand{\ee}{\end{equation}}
\newcommand{\bea}{\begin{eqnarray}}
\newcommand{\eea}{\end{eqnarray}}
\def\SS{{S}}

\begin{document}

\title{Defects, Rigid Holography and $C$-theorems}

\author{I. Carre\~no Bolla}
\author{Diego Rodriguez-Gomez}
\affiliation{Department of Physics, Universidad de Oviedo, \\ C/ Federico Garc\'ia Lorca  18, 33007  Oviedo, Spain}\affiliation{Instituto Universitario de Ciencias y Tecnolog\'ias Espaciales de Asturias (ICTEA),\\  C/~de la Independencia 13, 33004 Oviedo, Spain.}
\author{J.G. Russo}
\affiliation{Instituci\'o Catalana de Recerca i Estudis Avan\c{c}ats (ICREA), \\ Pg.~Lluis Companys, 23, 08010 Barcelona, Spain}\affiliation{Departament de F\' \i sica Cu\' antica i Astrof\'\i sica and Institut de Ci\`encies del Cosmos, \\ Universitat de Barcelona, Mart\'i Franqu\`es, 1, 08028 Barcelona, Spain }

\begin{abstract}
We consider  a general unitary scalar CFT with a linear defect in $D=4-\epsilon$ and a surface defect in $D=6-\epsilon$.
Using holography and the Hamilton-Jacobi formalism, 
we show that the $\beta$ functions controlling the defect RG flow are the gradient of the  entropy function. 
This  allows the proof that the relevant $C$-functions    decrease monotonically along the RG flow.
We provide evidence  that this property  also holds in the full quantum theory for general scalar field theories. An obstruction to the gradient property seems to appear at two loop order when  fermions are added.
\end{abstract}

\maketitle

\section{Introduction}

The Renormalization Group (RG) flow plays a central role in modern theoretical Physics, as it allows us to understand the relevant degrees of freedom of quantum systems at low energies. Heuristically, RG flows are expected to be irreversible. 
Refining this concept has been a major research stimulus over the last four decades, starting with the celebrated proof of the $C$-theorem in two dimensions \cite{Zamolodchikov:1986gt}. In the presence of defects --defined generically as operators supported on a $d$-dimensional submanifold $\mathcal{M}_d$ inside a $D$ dimensional bulk theory-- a similar picture is expected \cite{Jensen:2015swa,Kobayashi:2018lil,Giombi:2020rmc,Nishioka:2021uef,Wang:2021mdq,Cuomo:2021rkm,Shachar:2022fqk,Casini:2023kyj}. 
In this context, it has been realized that a relevant quantity to study defect RG flows is the free energy $\mathcal{F}_d$  when $\mathcal{M}_d=\mathbb{S}^d$,
defined as minus the logarithm of the partition function in the presence of a defect normalized by the partition function of the bulk theory without the defect. However, this quantity is scheme-dependent and is therefore not free from ambiguities. Denoting the radius of the $\mathbb{S}^d$ by $L$ and being $\Lambda $ a UV cut-off, $\mathcal{F}_d$ is given by the following expression for line and surface defects

\begin{equation}
\label{F}
    \mathcal{F}_d =\begin{cases} c^{(1)}\,(\Lambda L)-s_1 & d=1\,,\\ c^{(2)}(\Lambda L)^2+c^{(0)}-s_2\log(\Lambda L)& d=2\,.\end{cases}
\end{equation}
Only the entropy functions $s_d$ are universal, in the sense that they do not change under scaling of $\Lambda $.\footnote{
In $d=1$, the term $s$ has  a scheme dependence  since  it shifts by a  constant under a shift of $\Lambda$. However, this  is without relevance since one can define $s_1$ in scheme-independent way as 
$s_1=(1-L\partial_L) \mathcal{F}_1 $.} In fact, it has been argued that these coefficients are the 
pertinent $C$-functions for the defect RG flows in ambient Conformal Field Theories (CFT's): they are monotonically decreasing and, in particular, for flows interpolating between a UV and an IR fixed point their value decreases \cite{Cuomo:2021rkm,Casini:2023kyj} (see also  \cite{Giombi:2020rmc,Kobayashi:2018lil,Shachar:2022fqk}) and coincide with the appropriate central charges at the fixed points (in particular, in $d=2$, at fixed points, $s_2=\frac{b}{3}$).

Proving this has been a \textit{tour de force} over the years, where various approaches (information-theoretic, holographic as well as purely field-theoretic) have converged. In this note we will offer yet a new viewpoint on these monotonicity theorems. We will focus on 
line defects in $D=4$ and surface defects in $D=6$, where it is possible to consider a semiclassical limit that freezes the bulk quantum dynamics. As  will be shown below, in this limit it is possible to give a formula for $\mathcal{F}_d$ from which we obtain $s_d$. Moreover, we will show that, up to constant factors, the $\beta$ functions for the defect RG flow are the gradient of $s_d$, in particular proving a conjecture in \cite{Rodriguez-Gomez:2022gif,CarrenoBolla:2023sos}. Thus, a corollary of this formula is that, for bulk CFT's, the RG flow indeed minimizes $s_d$.

\section{Hamilton-Jacobi and the defect partition function}

We are  interested in the class of theories defined by the action
\begin{equation}
\label{accionuno}
    I=\int d^Dx\, \left[ \frac{1}{2}\big(\partial_\mu \phi_i\big)^2+V(\phi_i)\right]\,,
\end{equation}
where $V(\phi_i)$ is a generic homogeneous polynomial of  degree 
$n=3$ in $D=6$, and $n=4$ in $D=4$, that is,
$$
V^{6D}=\frac16\, \lambda_{ijk}\phi^i \phi^j \phi^k\ ,\quad 
V^{4D}=\frac1{24}\, \lambda_{ijkl}\phi^i \phi^j \phi^k \phi^l\ .
$$
The fields have classical dimension $d=\frac{D-2}{2}$, thus the classical theories are conformally invariant. 
In these theories we consider a trivial defect supported on a $d$-dimensional submanifold $\mathcal{M}_d$ (that is, $d=1$ in $D=4$ and $d=2$ in $D=6$).
Classically, the fields $\phi_i$ themselves are marginal deformations of the defect and, as such, must therefore be included with generic couplings $h_i$.

In \cite{Rodriguez-Gomez:2022gbz,Rodriguez-Gomez:2022gif,CarrenoBolla:2023sos} the 
theory was studied in the double-scaling limit where $h_i\to\infty$, $\lambda_{ijk}\to 0$
with fixed $h_i\lambda_{jkl}$ in $D=6$, and $\lambda_{ijkl}\to 0$
with fixed $h_ih_j\lambda_{klmn}$ in $D=4$.

This represents a semiclassical limit that freezes the quantum dynamics of the bulk theory. While bulk loops are suppressed, an infinite series of loops involving interactions with the defect remain. As a consequence, the bulk theory becomes a CFT coupled to the defect. In this limit the defect partition function is determined by the saddle point approximation and, consequently,

\begin{equation}
\label{I}
    \mathcal{F}_d=-\log \frac{Z_{\rm defect}}{Z_{\rm bulk}} =-\log \frac{\int \mathcal{D}\phi_i\,
      e^{- \SS }}{\int \mathcal{D}\phi_i\,e^{- I}}= -\log e^{-S}=S\, ,
      \end{equation} 
      with
     \begin{equation}
     \label{SS}
     \SS \equiv I-h^{(0)}_i\int_{\mathcal{M}_d}\phi_i\, .
\end{equation}
Here $h^{(0)}_i$ represent the bare couplings (we will reserve $h_i$ for the renormalized couplings), and $\SS $ is evaluated on the solution of the equations of motion for $\phi_i$ including the defect source. We can express this solution as an infinite series in powers of $h_i\lambda_{jkl}$ in $D=6$
and in powers of $h_ih_j\lambda_{klmn}$ in $D=4$. 

As  will be seen below, generically $S$ contains non-universal, scheme-dependent contributions, which  reflect the ambiguities described in the introduction. Considering only the universal pieces, we can read the entropy functions $s_d$ from $S_{\rm universal}$ as

\begin{equation}
   -S_{\rm universal}= \begin{cases} s_1\  &d=1\,,\\ s_2\log R\  &d=2\,.\end{cases}
\end{equation}

The renormalization procedure leads to $\beta_i$ functions for the $h_i$ couplings controlling the defect RG flow. This was studied in great detail in \cite{Rodriguez-Gomez:2022gbz,Rodriguez-Gomez:2022gif,CarrenoBolla:2023sos} for a linear defect
in 4D and for a  flat surface defect in 6D.
We now wish to study the properties of such flow, in particular explicitly computing  $s_d$ in the semiclassical limit for $\mathcal{M}_d=\mathbb{S}^d$. Our strategy will be to use rigid holography to evaluate $\SS $  in the solution of the classical equations of motion.
For this purpose, we perform a conformal transformation  from $\mathbb{R}^D$ to $\mathbb{H}_{d+1}\times \mathbb{S}^{d+1}$, so that the defect is located at the $\mathbb{S}^d$ boundary of $\mathbb{H}_{d+1}$, written in global coordinates as

\begin{equation}
ds^2=\frac{dr^2}{1+\frac{r^2}{L^2}}+r^2\,d\Omega_{d}^2+L^2\,d\Omega_{d+1}^2\,.
\end{equation}
For simplicity, in the following we will set $L=1$. Assuming that the field only has $r$-dependence, the bulk action becomes

\begin{equation}
\SS =\mathcal{C}_d \int_0^R dr \frac{r^{d}}{\sqrt{1+r^2}}\,\Big[ \frac{1}{2}\,(1+r^2)\, \big(\partial_r\phi_i\big)^2+V\Big]\,,
\end{equation}
where $\mathcal{C}_d={\rm Vol}(\mathbb{S}^d)\, {\rm Vol}(\mathbb{S}^{d+1})$. This action is UV divergent because the volume of $\mathbb{H}_{d+1}$ is infinite. We regulate this divergence by placing the  boundary of the hyperbolic space at a finite value $r=R$, which is assumed to be large ($R \rightarrow \infty$).
Note in particular that there is no mass term arising from the conformal coupling to curvature, by virtue of a cancellation between the contributions from $\mathbb{H}_{d+1}$ and $\mathbb{S}^{d+1}$.

In the holographic approach, the action $\SS $ can be computed by  solving the classical equations of motion on $\mathbb{H}_{d+1}$ with the boundary condition $\phi_i|_{\rm boundary}=(\frac{d}{4\pi})^{d}\,h_i$ (induced by the source term in the boundary), and then evaluating the action on the solution.
A more elegant approach is based on the Hamilton-Jacobi theory, where $r$ is interpreted as time. In what follows we shall use this approach\footnote{This approach to renormalization has a long history. For earlier applications of the Hamilton-Jacobi formalism in the context of holographic renormalization, see \cite{deBoer:1999tgo,Campos:2000yu,Martelli:2001tu,Martelli:2002sp}.}.

The canonical  momentum conjugated to $\phi_i$ is

\begin{equation}
\label{p}
p_i=\mathcal{C}_d\,\sqrt{1+r^2}\,r^{d}\,\partial_r\phi_i\,.
\end{equation}
Then, the Hamiltonian is

\begin{equation}
H=\frac{p_i^2}{2\,\mathcal{C}_d\,r^{d}\sqrt{1+r^2}}-\frac{\mathcal{C}_d\,r^{d}}{\sqrt{1+r^2}}\,V\,.
\end{equation}
This leads to the Hamilton-Jacobi equation

\begin{equation}
\frac{1}{2\mathcal{C}_d\,R^{d}\sqrt{1+R^2}}\,\Big(\frac{\partial \SS }{\partial \phi_i}\Big)^2-\frac{\mathcal{C}_d\,R^{d}}{\sqrt{1+R^2}}\,V=-\frac{\partial \SS } {\partial R}\,.
\end{equation}

Let us make the ansatz

\begin{equation}
\SS =\mathcal{C}_d\,v(R)\, \mathcal{H}(\phi_i)\, ,\quad v(R)\equiv \int_0^Rdr \frac{r^{d}}{\sqrt{1+r^2}}\,,
\end{equation}
where it is understood that $\phi_i$ is evaluated at the regularized boundary, located at large $R$ (representing a UV cutoff). 
Asymptotically, this leads to 

\begin{equation}
\frac{1}{2 d^2}\,\Big(\frac{\partial \mathcal{H}}{\partial \phi_i}\Big)^2-V=-\mathcal{H}\,.
\end{equation}
This is an equation for $\mathcal{H}(\phi)$, which is easily solved in perturbation theory in the $\lambda$ couplings. To the first few orders, we find

\begin{eqnarray}
\label{Huno}
\mathcal{H}&=&V-\frac{1}{2 d^2}\,V_i^2+\frac{1}{d^4}\,V_iV_j V_{ij} \\ \nonumber && -\frac{1}{8 d^6}\, (20V_{ij}V_{ik}V_jV_k+8V_i V_jV_kV_{ijk})+\cdots\,.
\end{eqnarray}
where $V_{i_1\cdots i_n}$ stands for $\frac{\partial}{\partial \phi_{i_1}}\cdots \frac{\partial}{\partial \phi_{i_n}}V$. After evaluating this expression on the boundary values $\phi_i|_{\rm boundary}=(\frac{d}{4\pi})^{d}\,h_i$ we get 

\begin{equation}
\label{Ssol}
\SS =\pi^{1-d}\,v(R)\,\mathcal{H}(h_i)\,,
\end{equation}

where now $\mathcal{H}(h_i)$ is

\begin{eqnarray}
\label{Hbeta}
    \mathcal{H}(h_i)&=&\Omega\,V-\frac{\Omega^2}{d^2}\,V_i^2+\frac{4\,\Omega^3}{d^4}\,V_{ij}V_iV_j\\ \nonumber && -\frac{8\,\Omega^4}{d^6}\,V_{ijk}V_iV_jV_k-\frac{20\,\Omega^4}{d^6}\,V_iV_{ij}V_{jk}V_k+\cdots \,.
\end{eqnarray}
where  $\Omega=\frac{d^2}{32 \pi^2}$ and $\mathcal{H}$ has been rescaled  by a numerical constant.

 We can now compute the $\beta$ function for the defect couplings $h_i$. In the Hamilton-Jacobi theory $p_i=\frac{\partial \SS }{\partial \phi_i}$. Thus, in the scheme
 naturally provided by holography, 

\begin{equation}
\label{betafinal}
\beta_i\equiv R\,\frac{\partial h_i}{\partial R}=\frac{2}{d}\,\frac{\partial \mathcal{H}}{\partial h_i} \, ,
\end{equation}
where we used \eqref{p} at $r=R$ and 
took the $R\to\infty $ limit.
Thus, in  this scheme, the $\beta$ function is indeed a gradient, in agreement  with similar   results obtained in \cite{Rodriguez-Gomez:2022gif,CarrenoBolla:2023sos}
for  planar defects. 
Furthermore, we now find that, for $\mathbb{S}^1$ and $\mathbb{S}^2$ defects, $\mathcal{H}$ represents the on-shell action, which we will now discuss.

\subsection{$C$-theorem}

Let us now comment on the implications for $C$-theorems.
These quantify the decrease in the low-energy effective  degrees of freedom of the system under the RG flow.

We first note that, as $R\to\infty$, the action \eqref{Ssol} inherits the following $R$ dependence

\begin{equation}
\label{divergences}
   v(R)=\begin{cases} R-1 +O(R^{-1}) \, ,  \  & {\rm for}\ d=1\ ,
   \\
   \frac{R^{2}}{2}-\frac{1}{2}\,\log (2e^{\frac{1}{2}}R) +O(R^{-2}) \, ,  \  & {\rm for}\ d=2\, .
    \end{cases}
\end{equation}
This is precisely the expected structure ({\it cf.} \eqref{F}). The divergences can be removed by adding suitable counterterms as in  standard holographic renormalization. For the present 4D and 6D theories we can add the following boundary counterterms
on the regularized boundary $\partial M|_R$:
\bea
&& d=2:\quad S_{\rm ct}^{6D}= \frac{\mathcal{H}}{8\pi^2}\int_{\partial M|_R} d\Sigma =\frac{1}{2\pi}\  R^2   \mathcal{H}\ ,
\nonumber\\
&& d=1: \quad S_{\rm ct}^{4D}= \frac{\mathcal{H}}{2\pi}\int_{\partial M|_R}  d\ell =  R   \mathcal{H}\ .
\eea
with $d\Sigma\equiv R^2\sin\theta d\theta d\varphi$ and $ d\ell\equiv R d\theta $.
Only the coefficient of $\log R$ in $d=2$ and the constant coefficient in $d=1$ are scheme-independent and thus physically meaningful. 
Therefore we obtain 

\begin{equation}
\label{Dd}
    s_1=\mathcal{H}\, ,\qquad s_2=\frac{\mathcal{H}}{2\pi}\ .
\end{equation}

In terms of the RG time $t=-\log R$, the relevant velocity vector in the space of couplings for the RG trajectories is nothing but 

\be
\beta_i =-\frac{\partial h_i}{\partial t}= 2\pi^{d-1}\, \frac{\partial s_d}{\partial h_i}\ . \label{eqt}
\ee
This relationship has several implications. To begin with, it implies that $s_d$ is an extremum at conformal points.
If the conformal point is a (UV) unstable fixed point, then away from this point $s_d$ is monotonically decreasing until it meets another (IR) stable fixed point.
More importantly, this relationship shows that $s_d$ must decrease along the time evolution of the RG flow. Indeed, 

\bea
\frac{d s_d}{d t} &=&
-\beta_i \frac{ \partial s_d}{\partial h_i}-\beta_{\lambda_{\alpha}} \frac{ \partial s_d}{\partial \lambda_{\alpha}}
\nonumber\\
&=& -\frac{\beta_i \beta_i}{2 \pi^{d-1}}-\beta_{\lambda_{\alpha}} \frac{ \partial s_d}{\partial \lambda_{\alpha}}\ ,
\eea
where $\lambda_\alpha$ denote the generic couplings $\{ \lambda_{ijkl}\} $.
For a conformal theory in the bulk we find
\be
\frac{d s_d}{d t}= -\frac{\beta_i \beta_i}{2 \pi^{d-1}} \leq 0\ .
\end{equation}
Therefore, the RG trajectories are such that $s_d$ is monotonically decreasing along the flow from the UV to IR, thus providing a  proof of the $C$-theorem for defect theories in the semiclassical limit. 
 The picture for the RG flow is similar to that of a particle rolling along a potential, where $\beta_i$ is the analogue of the velocity vector. 
 Indeed, if the particle is at a given (generic) point $\vec{h}(t_0)$  at time  $t=t_0$, at a later time $t=t_0+dt$ it will be at
\begin{equation}
\vec{h}(t_0+dt)= \vec{h}(t_0)-\vec{\beta}\,dt\ .
\end{equation}
Hence (assuming  $\beta_{\lambda_\alpha}=0$)
\be
s_d\big[ h(t_0+dt)\big] = s_d\big[h(t_0)\big] -\frac{|\vec \beta |^2}{2 \pi^{d-1}} d t \,. 
\ee
again exhibiting the fact that the entropy functions decrease along the RG flow.

\medskip

\subsection{The $\epsilon$ expansion}

Let us now consider the theories in $D=4-\epsilon$ and $D=6-\epsilon$. In our set-up, this means that we should now consider the theory in $\mathbb{H}_{d+1}\times \mathbb{S}^{d+1-\epsilon}$. This is implemented by shifting the dimension  in the sphere part of the geometry $d\Omega_{d+1}\rightarrow d\Omega_{d+1-\epsilon}$. This also induces a mass term coming from the conformal coupling to curvature which, to first order in $\epsilon$, amounts to the change in the potential 

\begin{equation}
V\rightarrow -\frac{d}{8}\,\epsilon\,\phi_i^2+V\,.
\end{equation}
When solving the HJ equation, the effect of the mass term is to induce a $\mathcal{O}(V^0)$ term in $\mathcal{H}$ proportional to $\epsilon$ (therefore also present in the free theory). It should be noted that this term  affects the equations for the higher orders in perturbation theory, but its effect
is $O(\epsilon^2)$ and can be neglected to leading order in $\epsilon$. Thus, the effect of the $\epsilon$ expansion is simply the expected shift in \eqref{Hbeta}

\begin{equation}
\mathcal{H}\rightarrow -\frac{d}{8}\epsilon\,h_i^2+\mathcal{H}\,.
\end{equation}

\section{Examples}

\subsection{Lines in a quartic theory in $D=4-\epsilon$}

For a generic quartic theory in  $D=4-\epsilon$, the above formulae \eqref{Hbeta}, \eqref{betafinal} 
yield

\begin{widetext}
\begin{eqnarray}
    s_1 &=& -\frac{\epsilon\,h^ih^i}{8}+ \frac{\lambda_{ijkl}h^ih^jh^kh^l}{768 \pi^2}  - \frac{\lambda_{iabc} \lambda_{iefg}  }{36864 \pi^4} h^a h^b h^c  h^e h^f h^g  + \frac{\lambda_{iabc} \lambda_{jefg} \lambda_{ijlm}   }{589824 \pi^6} h^a h^b h^c  h^e h^f h^g h^l h^m+ \cdots\,, \label{s1} \\
    \beta_i &=& -\frac{\epsilon\,h^i}{2}+ \frac{\lambda_{ijkl}}{96 \pi^2}\, h^jh^kh^l  - \frac{\lambda_{ijab} \lambda_{jcef}}{3072 \pi^4}\,  h^a h^b h^c  h^e h^f  +\frac{\lambda_{ijab}\lambda_{jkce} \lambda_{kfgl}}{49152 \pi^6} h^a h^b h^c h^e  h^f h^g h^l \nonumber \\ && + \frac{\lambda_{ijka}\lambda_{jbce} \lambda_{kfgl} }{147456\pi^6} h^a h^b h^c h^e  h^f h^g h^l +\cdots\,. \label{beta1}
\end{eqnarray}
\end{widetext}
(recall that $h^i$ represents the renormalized defect couplings). For simplicity in the presentation,  we omitted here the $\lambda^4$ term, which can be read from \eqref{Hbeta}, \eqref{betafinal}.

It is worth noting that the last two terms of \eqref{beta1} arise from differentiating a single term of $s_1$ in \eqref{s1}, and thus are related in a precise way. However, it should be remembered that some of the terms in the $\beta$-functions are scheme dependent from  three-loop onwards. In a generic scheme, the $\beta$ function coefficients of three-loop terms will not satisfy such relations and consequently the $\beta_i$ functions will not be given by a gradient. 
However, we have just seen that there is a natural scheme where the $\beta_i$ are given the simple gradient formula \eqref{betafinal} (a discussion comparing the different schemes can be found in  \cite{CarrenoBolla:2023sos}).

 The terms up to $O(\lambda^2)$ agree with the corresponding terms computed in (2.22) in  \cite{Pannell:2023pwz}
(note that here $h_i$ are defined with the reverse sign with respect to the definition in \cite{Pannell:2023pwz}).

For the familiar $V=\frac{\lambda}{4!}\phi^4$ potential, we find

\begin{eqnarray}
s_1&=& -\frac{\epsilon\,h^2}{8}+\frac{\lambda h^4}{768\pi^2}-\frac{\lambda^2h^6}{36864\pi^4}
\nonumber\\
&+&\frac{\lambda^3h^8}{589824\pi^6}-\frac{19 \lambda^4h^{10}}{113246208\pi^8}+\cdots\,,
\end{eqnarray}

\begin{eqnarray}
\beta &=& -\frac{\epsilon\,h}{2}+\frac{\lambda h^3}{96\pi^2}-\frac{\lambda^2h^5}{3072\pi^4}
\nonumber\\
&+&\frac{\lambda^3h^7}{36864\pi^6}-\frac{95 \lambda^4h^{9}}{28311552\pi^8}+\cdots\,.
\label{betacuar}
\end{eqnarray}
As a sanity check, the first two terms in $s$ exactly match those in \cite{Cuomo:2021kfm} (\textit{cf.} (3.26) in that reference), while the first three terms in $\beta$ exactly match the corresponding terms  in \cite{Cuomo:2021kfm} (\textit{cf.} (3.17) in that reference\footnote{Bulk loops start at order $\lambda^2$. These contributions are missing in the semiclassical formula \eqref{betacuar}, which  contains only the terms  with the highest power of $h$ for a given power of $\lambda $.}).

\subsection{Surfaces in a cubic theory in $D=6-\epsilon$}

We now consider a generic cubic theory  in $D=6-\epsilon$. In this case, the  formulae \eqref{Hbeta}, \eqref{betafinal} give

\begin{eqnarray}
\label{s2}
    2\pi s_2&=&  -\frac{\epsilon}{4} h^ih^i+ \frac{\lambda_{ijk} }{48\pi^2} h^ih^jh^k  - \frac{\lambda_{iab}\lambda_{icd} }{1024\pi^4} h^a h^b h^c h^d 
    \nonumber\\
    &+&
    \frac{\lambda_{iab}\lambda_{jcd} \lambda_{ije} }{8192\pi^6} h^a h^b h^c h^d h^e +\cdots\,,
\end{eqnarray}
    \begin{eqnarray}
    \label{betaseis}
    &&\beta_i = -\frac{\epsilon}{2} h^i+ \frac{\lambda_{ijk}}{16\pi^2} h^j h^k  - \frac{\lambda_{ijb}\lambda_{jcd}  }{256\pi^4} h^b h^c h^d 
    \\
       &&+\frac{\lambda_{ija}\lambda_{jkb}\lambda_{kcd} }{2048 \pi^6 } h^a h^b h^c h^d+ \frac{\lambda_{ijk}\lambda_{jab}\lambda_{kcd} }{8192 \pi^6 } h^a h^b h^c h^d+\cdots \nonumber \label{beta2}
\end{eqnarray}
As in the 4D case, the last two terms of  \eqref{betaseis} originate from a single term in \eqref{s2}, again showing that the gradient property imposes constraints on the coefficients of the $\beta $ functions.

Let us apply these formulas to a $O(N)$ model with  scalar fields $\sigma $ and $\phi_a$, $a=1,...,N$, and a potential
$$
V=\frac{\lambda_1}{2}\sigma \phi_a^2+\frac{\lambda_2}{6}\sigma^3\ .
$$ 
Let $h_{\sigma}$ and $h_{a}$ denote the renormalized defect couplings to $\sigma$ and $\phi_a$. Tuning $h_{a}=0$, we now find

\begin{equation}
2\pi s_2=-\frac{\epsilon}{4}h_{\sigma}^2+\frac{\lambda_2h_{\sigma}^3}{48\pi^2}-\frac{\lambda_2^2h_{\sigma}^4}{1024\pi^4}+\frac{\lambda_2^3h_{\sigma}^5}{8192\pi^6}-\frac{3\lambda_2^4h_{\sigma}^6}{131072\pi^8} +\cdots  
\nonumber
\end{equation}
\begin{equation}
\beta_{h_\sigma}=-\frac{\epsilon}{2}h_{\sigma}+\frac{\lambda_2h_{\sigma}^2}{16\pi^2}-\frac{\lambda_2^2h_{\sigma}^3}{256\pi^4}+\frac{5\lambda_2^3h_{\sigma}^4}{8192\pi^6}-\frac{9\lambda_2^4h_{\sigma}^5}{65536\pi^8} +\cdots \nonumber
\end{equation}
Note that these semiclassical contributions do not depend on $N$, as expected, as the $N$ dependence emerges only when incorporating bulk loops.
These expressions can be compared with various results in the literature.
Setting $\lambda_1=\lambda_2=0$, in the free theory we recover the result $s_2=-\frac{\epsilon}{8\pi}h_{\sigma}^2$ in \cite{Shachar:2022fqk,Cuomo:2023qvp}. Moreover, the first two terms in the $\beta$ function reproduce the results of \cite{Rodriguez-Gomez:2022gbz} (\textit{cf.} (64) in that reference).
At $O(\lambda^2)$ there are, in addition, bulk loop contributions recently computed by \cite{Giombi:2023dqs}.

Using \eqref{betaseis}, one can  also write the $\beta$ function  for  defect couplings $h_a\neq 0$.
The resulting formula reproduces the two loop results of \cite{Trepanier:2023tvb} (\textit{cf.} (4.4)), but in addition
provides all semiclassical contributions up to four loops.

\section{Beyond the semiclassical limit}

In the previous sections we have used  Hamilton-Jacobi theory combined with holography to compute $\mathcal{F}_d$ in a semiclassical limit. The holographic setup provides a natural scheme where the $\beta$ functions are a gradient of $s_d$, at least up to four loop orders. This ensures the monotonicity of $s_d$ along the RG flow. 
An important question is to what extent these considerations survive the inclusion of bulk loops. 

Let us first consider line operators  in scalar field theories in $D=4-\epsilon$. The $\beta $ functions for the general model \eqref{accionuno}
including bulk loops were computed in \cite{Pannell:2023pwz}
 up to $O(\lambda^2)$. Using the expression (2.22) for the $\beta_i$  in  \cite{Pannell:2023pwz}, it is easy to check that  $\partial_j\beta_i -\partial_j\beta_i =0$.
 Therefore, the $\beta$ functions are  still, remarkably, a gradient, now of a ``quantum-corrected" $\mathcal{H}^Q$ given by

\begin{equation}
\mathcal{H}^Q=\mathcal{H} +\frac1{24} V_{jkl} V_{jkl} -\frac14 V_{kl} V_{kl}\, ,
\end{equation}

However, it remains to check that  $\mathcal{H}^Q$ is proportional to $s_1$ (obtained from the quantum corrected $\mathcal{F}_1$). For the melonic tensor model in $D=4-\epsilon$ --which can be regarded as a particular quartic scalar theory-- both the beta functions and the $s_1$ were computed in \cite{Popov:2022nfq} in the large $N$ limit. In this limit it turns out that some contributions of bulk loops are equally relevant as the diagrams corresponding to the semiclassical limit (\textit{cf.} the diagram on the right of fig.7 in that reference). Interestingly, \cite{Popov:2022nfq}  finds that not only the $\beta$ function is a gradient, but also that the $\mathcal{H}^Q$ function is a multiple of $s_1$ (see (4.30) in \cite{Popov:2022nfq}).

Reference \cite{Pannell:2023pwz} also computes the $\beta$ function for line defects in $D=4-\epsilon$ in a model including fermions.
 Using the same notation as in \cite{Pannell:2023pwz}, we add to our action
\be
I_{\mathrm{fermion}}=I+ \int d^Dx\, \left(\frac12 y_{iab} \phi_i \psi_a \psi_b+ \mathrm{h.c.}\right)\,.
\ee
where $y_{iab}$ are symmetric in the fermion flavour indices. Defining $Y_{ij}= y_{iab} y^{*}_{jab}+y_{jab} y^{*}_{iab}$, 
and demanding that $\beta_i$ is a gradient, that is $\partial_j\beta_i -\partial_j\beta_i = 0$, we now get the condition
\be
\label{unio}
\lambda_{ik rs}Y_{kj}-\lambda_{jk rs}Y_{ki}=0\ ,
\ee
for all $r, s$. This implies that all matrices $\{ M^{(rs)}\}_{ij}= \lambda_{ij rs}$, with $r,s=1,...,N$ should commute with $Y_{ik}$.
For a general model,  this condition is not fulfilled and the $\beta_i$ functions do  not seem to be a gradient (at least in the scheme of \cite{Pannell:2023pwz}). 
For particular fermion couplings, one may have $Y_{ij}=Y_0\, \delta_{ij}$. In such a case the $\beta $ would still be a gradient.
An example is to take $y_{iab}=T^{i}_{ab}$, where 
$T^{i}$ are the generators of the Lie algebra of $O(N)$ in a symmetric representation. In a suitable basis, $Tr[T^iT^j]=a\delta_{ij}$, and \eqref{unio} is satisfied for any set of couplings $\{ \lambda_{ijkl}\}$. In this particular model with interacting scalar and fermion fields, the $\beta $ function is still of the form $\beta_i=\partial_i \hat H$. 

It should be noted that fermion contributions to $\beta_i$  appear only through Feynman diagrams that include fermion loops in the bulk.
 Therefore, fermion contributions are suppressed in the double scaling limit and do not alter the calculation of the defect coupling beta functions in the previous sections.

\section{Final comments}

In this note we have provided a proof of the monotonicity of RG flows for line defects in $D=4$ and surface defects in $D=6$ in general unitary CFT's with scalar fields in a semiclassical limit. In the case of {\it free} scalar field theories, the proof applies to the complete quantum theory. By mapping to $\mathbb{H}_{d+1}\times \mathbb{S}^{d+1}$  the problem of studying spherical defects becomes effectively one-dimensional, with the radial coordinate of $\mathbb{H}_{d+1}$  playing the role of time. Using the Hamilton-Jacobi formalism, we  derived the relevant formula for $s_d$ --here  explicitly written to four loop order-- which also governs the defect $\beta$  functions, given by the gradient of $s_d$ in the coupling space. Consequently, the renormalization group flow exhibits a resemblance to a particle's motion along a potential, thereby guaranteeing the monotonic decrease of the entropy functions along the flow. Importantly, this conclusion remains independent of the presence of an  IR defect fixed point.

Our analysis is conducted within the semiclassical limit, wherein the contributions of bulk loops are suppressed. In the holographic framework, this corresponds precisely to the large $N$ limit, which endows the bulk theory with classical behavior. Clearly, it is important to investigate whether the  structure found holds, to some extent, in the complete quantum theory. An initial step in this direction involves establishing whether the beta function in the full quantum theory can be expressed as the gradient of a coupling-dependent function within an appropriate scheme. As we have observed, this turns out to be the case for scalar field theories, at least up to the two-loop order. Subsequently, in a second step, the objective would be to establish the proportionality of this function to the defect entropy. At least in the melonic limit, we have found that the $\beta $ function indeed emerges as a gradient of the on-shell action. 
Although certain bulk diagrams are neglected in the melonic limit, strikingly this property  persists.
 On the other hand, upon the inclusion of fermions, the situation seems to be different, as even in the initial step, $\beta $ functions are not determined by a gradient in fermion models with generic couplings. It would be of great interest to delve into a more detailed examination of the underlying obstruction and explore its potential implications for $C$-theorems.

Our derivation of the $C$-theorems has been carried out specifically for lines in four dimensions and for surfaces in six dimensions.
It holds great interest to expand the current approach to encompass defects of arbitrary dimension $d$ in $D$-dimensional bulk theories, as well as to explore the implications of monodromy defects. In particular, a compelling direction lies in the generalization of $C$-theorems for $d=3$  and $d=4$. This would complement the recent findings  derived from the quantum information approach \cite{Casini:2023kyj}, offering further insights into the fundamental physical mechanisms that underlie these theorems.

\medskip

\section*{Acknowledgements}

D.R-G. thanks Maxime Tr\'epanier for useful conversations. J.G.R. acknowledges financial support from grants 2021-SGR-249 (Generalitat de Catalunya), MINECO  PID2019-105614GB-C21, and  from the State Agency for Research  through the ``Unit of Excellence María de Maeztu 2020-2023" (CEX2019-000918-M). The work of I.C.B. and D.R.G is partly supported by Spanish national grant MCIU-22-PID2021-123021NB-I00 as well as the Principado de Asturias grant SV-PA-21-AYUD/2021/52177.

\end{document}